\documentclass[aps,prl,twocolumn,superscriptaddress,amsfont,graphicx,nofootinbib,preprintnumbers]{revtex4}%
\usepackage{color,graphicx,epsfig}
\usepackage{ifpdf}
\usepackage{amsmath}
\usepackage{bm}
\usepackage{color}
\usepackage[english]{babel}
\usepackage{graphicx}%
\usepackage{amsfonts}%
\usepackage{amssymb}
\usepackage{braket}
\usepackage{hyperref}
\usepackage{enumerate}

\definecolor{nicered}{rgb}{0.7,0.1,0.1}
\definecolor{nicegreen}{rgb}{0.1,0.5,0.1}
\hypersetup{colorlinks,citecolor= nicegreen,linkcolor= nicered}

\newcommand{\beq}{\begin{equation}}
\newcommand{\eeq}{\end{equation}}
\newcommand{\bea}{\begin{eqnarray}}
\newcommand{\eea}{\end{eqnarray}}

\definecolor{Red}{rgb}{1.,0.,0.}

\def\OMIT#1{}

\begin{document}

\def\Argonne{High Energy Physics Division, Argonne National Laboratory, Argonne, IL 60439, USA}
\def\Northwestern{Department of Physics \& Astronomy, Northwestern University, Evanston, IL 60208, USA}

% \preprint{}

\author{Simone Alioli}
\email[Electronic address: ]{simone.alioli@unimib.it}
\affiliation{Universit\'a degli Studi di Milano-Bicocca \& INFN, Piazza della Scienza 3, Milano 20126, Italia}

\author{Radja Boughezal}
\email[Electronic address: ]{rboughezal@anl.gov}
\affiliation{\Argonne}

\author{Emanuele Mereghetti}
\email[Electronic address: ]{emereghetti@lanl.gov}
\affiliation{Theoretical Division, Los Alamos National Laboratory, Los Alamos, NM 87545, USA}

\author{Frank Petriello}
\email[Electronic address: ]{f-petriello@northwestern.edu}
\affiliation{\Argonne}
\affiliation{\Northwestern}

\title{Novel angular dependence in Drell-Yan lepton production via
  dimension-8 operators}

\date{\today}
\begin{abstract}

We study the effects of dimension-8 operators on Drell-Yan production
of lepton pairs at the Large Hadron Collider (LHC).  We
identify a class of operators that leads to novel angular dependence
not accounted for in current analyses.  The observation of such
effects would be a smoking-gun signature of new physics appearing at
the dimension-8 level.  We propose an extension of the currently used
angular basis and show that these effects should be
observable in future LHC analyses for realistic values of the
associated dimension-8 Wilson coefficients.

\end{abstract}

\maketitle
%%%%%

%%%%%%%%%%%%%%%%%%%%%%%%%%%%%%%%%%%%%%%%%%%
\section{Introduction} \label{sec:intro}
%%%%%%%%%%%%%%%%%%%%%%%%%%%%%%%%%%%%%%%%%%%

The Standard Model (SM) has so far been remarkably successful in
describing all data coming from the Large Hadron Collider (LHC) and
elsewhere. Although the search for new particles beyond those predicted
in the SM  will continue at the high-luminosity LHC, it is becoming
increasingly important to search for potentially small and subtle
indirect signatures of new physics.  A convenient theoretical
framework for performing such searches when only the SM particles are
known is the SM effective field theory (SMEFT) which contains higher-dimensional operators
formed only from SM fields.   The SMEFT is an expansion in an energy scale $\Lambda$ at
which the effective theory breaks down and new fields must be added to
the Lagrangian.  The leading dimension-6 operators
characterizing deviations from the SM have been
classified~\cite{Buchmuller:1985jz, Grzadkowski:2010es} (there is a
dimension-5 operator that violates lepton number \cite{Weinberg:1979sa}, which does not play a role in our discussion).

Less is known about terms at dimension-8 and beyond in the SMEFT
expansion.  The number of operators at each order in the expansion has
been determined~\cite{Henning:2015alf}, and initial ideas on how to systematically derive the
structure of these operators have appeared~\cite{Hays:2018zze}.  Some
phenomenological consequences of dimension-8 operators in the
SMEFT have been studied~\cite{Degrande:2013kka,Hays:2018zze}.  Although
their effects are usually suppressed with respect to dimension-6
operators, dimension-8 terms are sometimes the leading contributions
to observables due to symmetry considerations or the structure of the
corresponding SM amplitudes~\cite{Azatov:2016sqh}.  In such cases
it is important to quantify their effects in order to guide
experimental searches.

In this note we point out that a class of dimension-8 operators in the
SMEFT generate novel angular dependences in Drell-Yan lepton-pair
production not accounted for in current experimental
analyses~\cite{Chatrchyan:2011ig,ATLAS:2012au,Aad:2016izn,Khachatryan:2015paa}.
They are not generated at leading-order by dimension-6 operators in
the SMEFT, nor by QCD effects in the SM.  They are only generated in
the SM by higher-order electroweak corrections, which we demonstrate
here to be small.  This offers the possibility of extending the
current experimental studies to search for this potential smoking-gun
signature of new physics appearing through dimension-8 effects.  We
note that such dimension-8 operators could be generated in an
ultraviolet completion by vector
leptoquarks, which would also induce dimension-6 effects~\cite{Dorsner:2016wpm}.  They could
also be generated without dimension-6 contributions by massive spin-two particles~\cite{Falkowski:2016glr}.

The typical angular analysis of lepton-pair production through either
charged or neutral currents proceeds by expanding the
differential cross section in terms of spherical harmonics:
\begin{eqnarray}
\frac{d \sigma}{dm_{ll}^2 dy d\Omega_l} &=& \frac{3}{16\pi} \frac{d
  \sigma}{dm_{ll}^2 dy}
\left\{(1+c_{\theta}^2)+\frac{A_0}{2}(1-3c_{\theta}^2) \right. \nonumber \\
 && \left. +A_1
s_{2\theta}c_{\phi}+\frac{A_2}{2}s^2_{\theta}c_{2\phi} +A_3
                                         s_{\theta}c_{\phi}+A_4 c_{\theta}\right. \nonumber \\
 && \left.  +A_5 s^2_{\theta}s_{2\phi} +A_6 s_{2\theta}s_{\phi} +A_7
    s_{\theta}s_{\phi}
     \right\}.
     \label{eq:oldCSexp}
\end{eqnarray}
Here, $m_{ll}$ is the invariant mass of the lepton system, $y$ is the
rapidity of the $W$ or $Z$-boson that produces the lepton pair, and
$\Omega_l$ is the solid angle of a final-state lepton. The lepton
angles are typically defined in the Collins-Soper
frame~\cite{Collins:1977iv} and we have used the notation $s_{\alpha}$
and $c_{\alpha}$ to represent their sine and cosine, respectively.  In
the SM, the leptons are produced by an $s$-channel spin-one current,
so in the squared amplitude spherical harmonics up to $l=2$ are
allowed.  We show that certain two-derivative dimension-8 operators in
the SMEFT populate the $l=2$ partial wave at the amplitude level,
allowing for $l=3$ spherical harmonics in the angular expansion when
interfered with the SM amplitude.  Dimension-6 operators cannot
generate $l=2$ partial waves at the amplitude level, making their
appearance a hallmark of the dimension-8 SMEFT.  Searching for such
effects requires extending the usual angular analysis as we
demonstrate later in Eq.~(\ref{eq:newCSexp}).

Our paper is organized as follows.  We first review the operator basis for
SMEFT, focusing on operators relevant for lepton pair
production at dimension-6 and dimension-8.  We consider operators relevant for
both leading-order (LO) and next-to-leading-order (NLO) in the QCD
coupling constant.  We then present
formulae for LO production and demonstrate the need to
expand the usual spherical harmonic basis.  Finally, we present numerical results
for neutral-current production at the LHC, where we also show that
the predicted SM results for these angular dependences arising from
higher-order electroweak corrections are small.

%%%%%%%%%%%%%%%%%%%%%%%%%%%%%%%%%%%%%%%%%%%
\section{Review of the SMEFT} \label{sec:smeft}
%%%%%%%%%%%%%%%%%%%%%%%%%%%%%%%%%%%%%%%%%%%

We review in this section aspects of the SMEFT relevant for
our analysis of the angular dependence of lepton-pair production.  The SMEFT is an extension of the SM Lagrangian to include terms
suppressed by an energy scale $\Lambda$ at which the ultraviolet completion
becomes important.  Truncating the expansion in $1/\Lambda$ at dimension-8, and
ignoring operators of odd-dimension which violate lepton number, we
have
\begin{equation}
{\cal L} = {\cal L}_{SM}+\frac{1}{\Lambda^2} \sum_i C_{6,i} {\cal
  O}_{6,_i} + \frac{1}{\Lambda^4} \sum_i C_{8,i} {\cal
  O}_{8,i}.
\end{equation}
Operators of dimension-6 have been extensively studied in the
literature \cite{Cirigliano:2012ab,Brivio:2017btx,Alioli:2017ces,Alioli:2017nzr,Alioli:2018ljm,Dawson:2019xfp,Carrazza:2019sec}.
The overall electroweak couplings that govern lepton-pair production are shifted in SMEFT.  Since these clearly lead to only an overall shift
of the couplings and not to any new angular terms we do not explicitly
consider them here.
In addition,  Drell-Yan lepton-pair production receives contributions from
several classes of dimension-6 operators that affect angular distributions.
Two types of operators have non-vanishing interference with the SM, and lead to genuine dimension-6 effects in the cross sections.
In the notation of Ref.~\cite{Grzadkowski:2010es,Dedes:2017zog}, these
belong to the classes
\begin{itemize}
 \item $\psi^2 \varphi^2 D$: these include operators with a single
   derivative and a fermion bilinear of the form
   \begin{equation}\label{eq:1}
   {\cal O}_{6,\varphi e} = (\varphi^{\dagger} i
   \overleftrightarrow{D}_{\mu} \varphi) ( \bar{e} \gamma^{\mu} e),
   \end{equation}
   where $\varphi$ denotes the Higgs doublet, $e$ a right-handed
   lepton singlet, $D_\mu$ a covariant derivative, and $\overleftrightarrow{D}_{\mu}= \overrightarrow{D}_\mu - \overleftarrow{D}_\mu$.
   Operators of this form simply shift the SM coupling of the fermions
   to gauge bosons. In charged-current processes, these interactions involve purely left-handed quarks and leptons
   and lead to exactly the same angular dependence as in the SM.  For
   neutral currents, operators in this class might shift
   the relative importance of left- and right-handed couplings with
   respect to the SM, and could manifest themselves in high-precision measurements
   of angular coefficients such as $A_4$.

 \item $\psi^4$: four-fermion operators with the same chiral structure as the SM, such as
   \begin{equation}
    {\cal O}_{6,eu} = (\bar{e} \gamma^{\mu}
    e)(\bar{u} \gamma_{\mu} u),
    \end{equation}
    where $u$ denotes a right-handed up-quark field. These operators
   have been extensively studied.  It is straightforward
   to see that these produce the same lepton angular dependences as in
   the SM, as they can be obtained by integrating out new spin-one
   $W^{\prime}$ or $Z^{\prime}$ gauge bosons.

\end{itemize}

In addition, the dimension-6 SMEFT Lagrangian contains several more operators that do not interfere with the SM,
and thus contribute to the cross section at $\mathcal
O(v^4/\Lambda^4)$. They belong to the following classes.
\begin{itemize}

\item $\psi^2 X \varphi$: these include dipole operators coupled to gauge
  fields such as
  \begin{equation}
   {\cal O}_{6,eW} = (\bar{l}\sigma^{\mu\nu} e)\tau^I \varphi W^I_{\mu\nu},
   \end{equation}
   where $l$ denotes a left-handed lepton doublet and $\tau^I$ an $SU(2)$ Pauli matrix,
   and similar operators involving quarks and the $U(1)_Y$ gauge boson.

   \item $\psi^2 \varphi^2 D$: in addition to the operators considered before, one can introduce the right-handed charged-current operator
   \begin{equation}
   {\cal O}_{6,\varphi u d} = (\tilde{\varphi}^{\dagger} i
   {D}_{\mu} \varphi) ( \bar{u} \gamma^{\mu} d) + {\rm h.c.},
   \end{equation}
where $u$ and $d$ are right-handed quark fields.

 \item $\psi^4$: four-fermion operators with chiral structure different from the SM, such as the scalar operator
   \begin{equation}
    {\cal O}_{6,ledq} = \bar{l}^i  e \,\bar d  q^i,
    \end{equation}
    where $q$ is a left-handed quark doublet.
\end{itemize}
As discussed in Ref. \cite{Alioli:2018ljm},
these operators can induce dramatic deviations from the SM expectations in the $A_i$ coefficients, especially
at large dilepton invariant masses. However, they do not generate any new angular dependence
and their effect is fully captured by Eq.~\eqref{eq:oldCSexp}.
This statement remains true upon including QCD corrections, since these diagrammatic
contributions feature a gluon connecting the two initial-state quarks
and do not affect the spin-one (or spin-zero) current that produces the lepton
pair. Only an electroweak correction where a gauge boson connects
an initial-state quark to a final-state lepton can populate a $l > 1$
partial wave.  We discuss this possibility in the case of the
higher-order SM corrections later in this note.

At dimension-8 a larger variety of operator classes can contribute.
We use the {\tt HSMethod} code~\cite{Henning:2015alf} to obtain the correct number of
operators with a given field content.  We note that many of the
operators relevant to our study were previously considered in
Ref.~\cite{Hays:2018zze}.  We have confirmed the number and structure
of the operators found there.
\begin{itemize}

\item $\psi^2 \varphi^4 D$: this category has been studied in
  Ref.~\cite{Hays:2018zze} and contains operators such as
  \begin{equation}
{\cal O}_{8,q1} = i (\bar{q} \gamma^{\mu} q) (\varphi^{\dagger}
 \overleftrightarrow{D}_{\mu}\varphi) (\varphi^{\dagger} \varphi).
    \end{equation}
    These clearly lead to shifts in the fermion-gauge boson vertices and no
    new kinematic effects, as confirmed by explicit calculation in
    Ref.~\cite{Hays:2018zze}.

  \item  $\psi^2 \varphi^2 D^3$: these include operators of the form
    \begin{equation}
     {\cal O}_{8,3q1} = i (\bar{q} \gamma^{\mu} D^{\nu}q)
     (D^2_{(\mu\nu)} \varphi^{\dagger} \varphi).
   \end{equation}
      These only shift the fermion-gauge boson vertices, as confirmed
      in Ref.~\cite{Hays:2018zze}.

\item $\psi^4 \varphi^2$: these include four-fermion operators such
  as
  \begin{equation}
{\cal O}_{8,eu}=(\bar{e} \gamma^{\mu}
    e)(\bar{u} \gamma_{\mu} u) (\varphi^{\dagger}\varphi).
   \end{equation}
   These clearly shift the dimension-6 couplings leading to the same
   angular dependence as before.  The remaining operators relevant for
   lepton-pair production can be obtained by considering both fermion
   doublets and singlets, and by judicious insertions of the Pauli matrices $\tau^I$.

\item $\psi^4 D^2$: we begin by considering operators with left-handed
  fermion doublets only.  There are four such operators, which we
  write in the following way:
  \begin{eqnarray}
   {\cal O}_{8,lq\partial 1} &=& (\bar{l} \gamma_{\mu} l) \partial^2
                                 (\bar{q} \gamma^{\mu} q) ,\nonumber
    \\
     {\cal O}_{8,lq\partial 2} &=& (\bar{l} \tau^I \gamma_{\mu} l) \partial^2
                                 (\bar{q} \tau^I \gamma^{\mu} q)
                                   ,\nonumber \\
      {\cal O}_{8,lq\partial 3} &=& (\bar{l} \gamma_{\mu} \overleftrightarrow{D}_{\nu} l)
                                 (\bar{q} \gamma^{\mu} \overleftrightarrow{D}^{\nu} q)
                                    ,\nonumber \\
      {\cal O}_{8,lq\partial 4} &=& (\bar{l}\tau^I \gamma_{\mu} \overleftrightarrow{D}_{\nu} l)
                                 (\bar{q}\tau^I \gamma^{\mu}
                                    \overleftrightarrow{D}^{\nu} q).
   \label{eq:dim8d2ops}
   \end{eqnarray}
   The operators $ {\cal O}_{8,lq\partial 1}$ and $ {\cal
     O}_{8,lq\partial 2}$ lead only to an energy-dependent shift of
   the dimension-6 four-fermion couplings.  This is clear from their
   form and can also be confirmed by explicit calculation.  The
   remaining two operators are more interesting.  Considering the
   lepton bilinears present in ${\cal O}_{8,lq\partial 3}$ and ${\cal
     O}_{8,lq\partial 4}$, we see that they each contain two free
   Lorentz indices $\mu$ and $\nu$.  This implies that they can couple to a spin-two
   current, which can be represented as a two-index polarization
   tensor $\epsilon_{\mu\nu}$.  The amplitude therefore
   contains a new $l=2$ partial wave not present in
   previous contributions.  We confirm this later by explicit
   calculation.  For charged-current production only $ {\cal
     O}_{8,lq\partial 4}$ would contribute.

   We now extend our basis of operators to include right-handed
   fermion fields as well, and focus on operators containing the
   $\gamma^{\nu}\overleftrightarrow{D}^{\mu}$ structure necessary for the
   angular dependence of interest.  We find an additional five
   operators:
   \begin{eqnarray}
 {\cal O}_{8,ed\partial 2} &=& ( \bar{e}  \gamma_{\mu} \overleftrightarrow{D}_{\nu} e)( \bar{d}
 \gamma^{\mu} \overleftrightarrow{D}^{\nu} d),  \nonumber
   \\
 {\cal O}_{8,eu\partial 2} &=& ( \bar{e}  \gamma_{\mu} \overleftrightarrow{D}_{\nu} e)( \bar{u}
 \gamma^{\mu} \overleftrightarrow{D}^{\nu} u), \nonumber
     \\
{\cal O}_{8,ld\partial 2} &=&( \bar{l}  \gamma_{\mu} \overleftrightarrow{D}_{\nu} l)( \bar{d}
 \gamma^{\mu} \overleftrightarrow{D}^{\nu} d),  \nonumber
   \\
{\cal O}_{8,lu\partial 2} &=&( \bar{l}  \gamma_{\mu} \overleftrightarrow{D}_{\nu} l)( \bar{u}
 \gamma^{\mu} \overleftrightarrow{D}^{\nu} u),  \nonumber
   \\
{\cal O}_{8,qe\partial 2} &=& ( \bar{e}  \gamma_{\mu} \overleftrightarrow{D}_{\nu} e)( \bar{q}
                               \gamma^{\mu} \overleftrightarrow{D}^{\nu} q).
   \end{eqnarray}
   We arrive at the following seven operators that can
   contribute to $l=2$ partial waves for the neutral-current
   amplitude: $ {\cal O}_{8,lq\partial 3}$, $ {\cal O}_{8,lq\partial
     4}$,  $ {\cal O}_{8,eu\partial 2}$, $ {\cal O}_{8,ed\partial
     2}$,  $ {\cal O}_{8,lu\partial 2}$, $ {\cal O}_{8,ld\partial
     2}$ and $ {\cal O}_{8,qe\partial 2}$.

 \end{itemize}

 We next discuss the dimension-8 operators containing gluons that can
 contribute to Drell-Yan lepton-pair production at NLO in the QCD
 coupling constant.  As we find that none of these operators
 contribute to the angular dependence that is the major point of this
 note, we discuss them briefly for left-handed doublets only.

 \begin{itemize}

  \item $\psi^4 G$: there are four such operators that contribute at
    dimension-8 for left-handed fermion fields.  We list the two
    distinct operator structures that appear below,  the remaining two can
    be obtained by changing the gluon field-strength tensor to the
    dual one:
    \begin{eqnarray}
      {\cal O}_{8,lqG1} &=& (\bar{l} \gamma^{\mu}
                          l)(\bar{q} t^A \gamma_{\nu} q) G^A_{\mu\nu},\nonumber \\
        {\cal O}_{8,lqG2} &=& (\bar{l} \tau^I\gamma^{\mu}
                          l)(\bar{q} \tau^I   t^A \gamma_{\nu} q) G^A_{\mu\nu}.
    \end{eqnarray}
    The lepton bilinears in these operators couple to a spin-one
    current, indicating that they lead to the usual angular
    dependence found in the SM.  We have confirmed this by explicit calculation.

  \item $\psi^2 \varphi^2 DG$: these are corrections to the quark
    bilinear that also contain a gluon field.  Specializing to
    left-handed quarks we find eight such operators.  We list the four
    distinct operator structures that appear, the remaining four can
    be obtained by changing the gluon field-strength tensor to the
    dual one:
    \begin{eqnarray}
    {\cal O}_{8,qG1} &=& (\bar{q} t^A \gamma^{\nu} q)
                       \partial^{\mu}(\varphi^{\dagger}\varphi ) G^A_{\mu\nu}, \nonumber \\
         {\cal O}_{8,qG2} &=& (\bar{q} t^A \gamma^{\nu} q)
                            (\varphi^{\dagger} i
                            \overleftrightarrow{D}^{\mu} \varphi)
                            G^A_{\mu\nu},  \nonumber \\
       {\cal O}_{8,qG3} &=& (\bar{q} \tau^I t^A \gamma^{\nu} q)
                       D^{\mu}(\varphi^{\dagger} \tau^I \varphi )
                          G^A_{\mu\nu}, \nonumber \\
       {\cal O}_{8,qG4} &=& (\bar{q} \tau^I t^A \gamma^{\nu} q)
                            (\varphi^{\dagger} \tau^I i
                            \overleftrightarrow{D}^{\mu} \varphi)
                            G^A_{\mu\nu}.
    \end{eqnarray}
    The operator $ {\cal O}_{8,qG1}$ requires a physical Higgs boson and
    therefore does not contribute to dilepton production.   We have
    checked by explicit calculation that $ {\cal O}_{8,qG3} $
    contributes in the same way as operators in the $\psi^2 \varphi^2
    D^3$ category, while $ {\cal O}_{8,qG2}$ and $ {\cal O}_{8,qG4}$ give similar
    contributions as $ {\cal O}_{8,lqG1}$ and $ {\cal O}_{8,lqG2} $.  None
    of these operators introduces novel
    angular dependence.

 \item $\psi^2 D X G$:
 these induce local interactions between two quarks, a weak boson and a gluon.  We find eight operators
 with left-handed quarks that contribute to Drell-Yan at NLO.  We list the two
    distinct operator structures that appear, the remaining can
    be obtained by changing the gluon field-strength tensor to the
    dual one, and by replacing the $SU(2)_L$ with the $U(1)_Y$ field strength.
    \begin{eqnarray}
    {\cal O}_{8,qWG1} &=& \left( \bar{q} t^A \tau^{I} \gamma^{(\mu}  i \overleftrightarrow D^{\nu)}) q \right) W^I_{\mu\rho} G^{A\, \rho}_\nu, \nonumber \\
    {\cal O}_{8,qWG2} &=& \left( \bar{q} t^A \tau^{I} \gamma^\mu q \right) \left( W^I_{\alpha\beta} \overleftrightarrow D_{\mu} G^{A\, \alpha\beta} \right).
    \end{eqnarray}
 Here $\gamma^{(\mu}  \overleftrightarrow D^{\nu)} = (\gamma^{\mu}   \overleftrightarrow D^{\nu} + \gamma^{\nu}   \overleftrightarrow D^{\mu})/2$.
 In this case, the leptons arise from the decay of a spin-one weak
 boson, and thus the angular distributions are described by
 Eq. \eqref{eq:oldCSexp}. We have verified this by an explicit calculation.

 \end{itemize}

%%%%%%%%%%%%%%%%%%%%%%%%%%%%%%%%%%%%%%%%%%%
\section{Angular dependence with dimension-8 effects} \label{sec:angdep}
%%%%%%%%%%%%%%%%%%%%%%%%%%%%%%%%%%%%%%%%%%%

It is straightforward to calculate the matrix elements for the
LO partonic process $u(p_1)\bar{u}(p_2) \to l(p_3)\bar{l}(p_4)$ given the operators in
the previous section.  We focus on ${\cal O}_{8,lq \partial 3}$ in
Eq.~(\ref{eq:dim8d2ops}) as an example.  Keeping only the leading
interference of this operator with the SM contribution, we find the following
SMEFT-induced correction to the matrix-element squared:
\begin{eqnarray}
\Delta |{\cal M}_{u\bar{u}}|^2 &=& -\frac{C_{8,lq\partial 3}}{\Lambda^4}
\,{\hat c}_{\theta}(1+\hat c_{\theta})^2 \frac{\hat{s}^2}{6} \times \nonumber \\ & & \left[e^2 Q_u
  Q_e  +\frac{g^2 g_L^u g_L^e \hat{s}}{c_W^2(\hat{s}-M_Z^2)} \right].
\label{eq:CLLd3amp}
\end{eqnarray}
Here, $\hat{s}$ denotes the usual partonic Mandelstam invariant
$\hat{s} = (p_1+p_2)^2$, $g$ is the $SU(2)$ coupling constant, $c_{W}$
is the cosine of the weak mixing angle, $e$ is the $U(1)_{EM}$
coupling constant, $Q_i$ is the charge of fermion $i$, $g_L^i$ are
the left-handed couplings to the $Z$-boson following the notation of Ref.~\cite{Denner:1991kt}.
$C_{8,lq\partial 3}$ is the Wilson coefficient associated with the
operator under consideration, and $\hat c_{\theta}$ is the angle between
the beam direction and the outgoing lepton direction.  At LO, the cosine of the polar angle $c_{\theta}$  in the Collins-Soper frame used
in the LHC analyses of Refs. \cite{Aad:2016izn,Khachatryan:2015paa}
is related to $\hat c_\theta$ by  $c_{\theta}= \pm \hat c_{\theta}$,
with positive (negative) sign if the longitudinal momentum of the
dilepton pair is along (opposite) to the beam direction.  We note that
the amplitude for $\bar{u}(p_1)u(p_2) \to l(p_3)\bar{l}(p_4)$ can be
obtained by taking $\hat c_{\theta} \to - \hat c_{\theta}$.  The down-quark
channel can be obtained by appropriate changes in the SM couplings.

This contribution to the differential cross section contains a
$c_{\theta}^3$ dependence that cannot be described by
Eq.~(\ref{eq:oldCSexp}).  The reason for this was given in the
previous section when discussing the operators of
Eq.~(\ref{eq:dim8d2ops}): the traditional formulation of the
Collins-Soper moments assumes that the lepton pair is produced in the
$s$-channel by a spin-one current, which is not the case for $ {\cal
  O}_{8,lq\partial 3}$.  Only the seven dimension-8 operators in the
$\psi^4 D^2$ category identified in the
previous section lead to an angular dependence not already described by
Eq.~(\ref{eq:oldCSexp}).

In order to account for this new signature of dimension-8 effects we
propose extending the parameterization of Eq.~(\ref{eq:oldCSexp}) to
the following:
\begin{eqnarray}
\frac{d \sigma}{dm_{ll}^2 dy d\Omega_l} &=& \frac{3}{16\pi} \frac{d
  \sigma}{dm_{ll}^2 dy}
\left\{(1+c_{\theta}^2)+\frac{A_0}{2}(1-3c_{\theta}^2) \right. \nonumber \\
 && \left. +A_1
s_{2\theta}c_{\phi}+\frac{A_2}{2}s^2_{\theta}c_{2\phi} +A_3
                                         s_{\theta}c_{\phi}+A_4 c_{\theta}\right. \nonumber \\
 && +A_5 s^2_{\theta}s_{2\phi} +A_6 s_{2\theta}s_{\phi} +A_7
    s_{\theta}s_{\phi}     \nonumber \\
 &&+ B_3^e s_{\theta}^3
     c_{\phi}  + B_3^o s_{\theta}^3s_{\phi} +B_2^e s_{\theta}^2
     c_{\theta} c_{2\phi}  \nonumber \\
 && +B_2^o s_{\theta}^2
     c_{\theta} s_{2\phi} +\frac{B_1^e }{2}
    s_{\theta}(5c_{\theta}^2-1) c_{\phi}  \\ \nonumber
  && \left.
  +  \frac{B_1^o }{2} s_{\theta}(5c_{\theta}^2-1) s_{\phi}  + \frac{B_0}{2} (5 c_{\theta}^3-3c_{\theta})
     \right\}.
     \label{eq:newCSexp}
\end{eqnarray}
We have used the combinations of spherical harmonics
\begin{equation}
Y_3^0, \;\;Y_3^{1} \pm Y_3^{-1},
\;\; Y_3^{2} \pm Y_3^{-2},\;\; Y_3^{3} \pm Y_3^{-3}.
\end{equation}
in forming the basis for the new $B_i^{e,o}$ coefficients.  The superscripts $e,o$ on the new $B_i$ coefficients refer to either
even or odd under T-reversal~\cite{Hagiwara:1984hi}.  The
amplitude of Eq.~(\ref{eq:CLLd3amp}) populates the $B_0$ coefficient.
The $B_i^{o,e}$ coefficients with $i>0$ are first populated at ${\cal
  O}(\alpha_s)$.

%%%%%%%%%%%%%%%%%%%%%%%%%%%%%%%%%%%%%%%%%%%
\section{Numerical results} \label{sec:numerics}
%%%%%%%%%%%%%%%%%%%%%%%%%%%%%%%%%%%%%%%%%%%

We present here numerical results for neutral-current lepton-pair production at the LHC
to assess the potential observation of these effects.  We assume
$\sqrt{s}=14$ TeV collisions.  Our hadronic results use the NNPDF 3.1
parton distribution functions extracted to NLO precision~\cite{Ball:2017nwa}, and
assume an on-shell electroweak scheme with $G_{\mu}$, $M_W$, and
$M_Z$ taken as input parameters.  Since we are interested in higher-dimensional operators that grow with
energy we impose the following cut on the invariant mass of the
final-state system: $m_{ll}>100$ GeV.  Only $B_0$ is
generated at this leading order in QCD perturbation theory, so we focus on this
coefficient here.  We set the renormalization and
factorization scales to $\mu = m_{ll}$.

As mentioned earlier, while the $B_i$ are not generated in the SM from
perturbative QCD corrections, they can be obtained from higher-order
electroweak effects.  The leading contributions to the $B_0$
coefficient are the angular-dependent next-to-leading logarithmic (NLL)
electroweak Sudakov logarithms (the higher $B_i$ coefficients require
a mixed ${\cal O}(\alpha\alpha_s)$ perturbative correction which we do
not consider).  The leading logarithms depend only on
the Mandelstam invariant $\hat{s}$, and therefore do not induce any
$B_i$ coefficients.  We study the leading one-loop NLL
electroweak Sudakov logarithms in the SM using the results of
Ref.~\cite{Denner:2006jr}.

\begin{figure}[h!]
\centering
\includegraphics[width=0.5\textwidth]{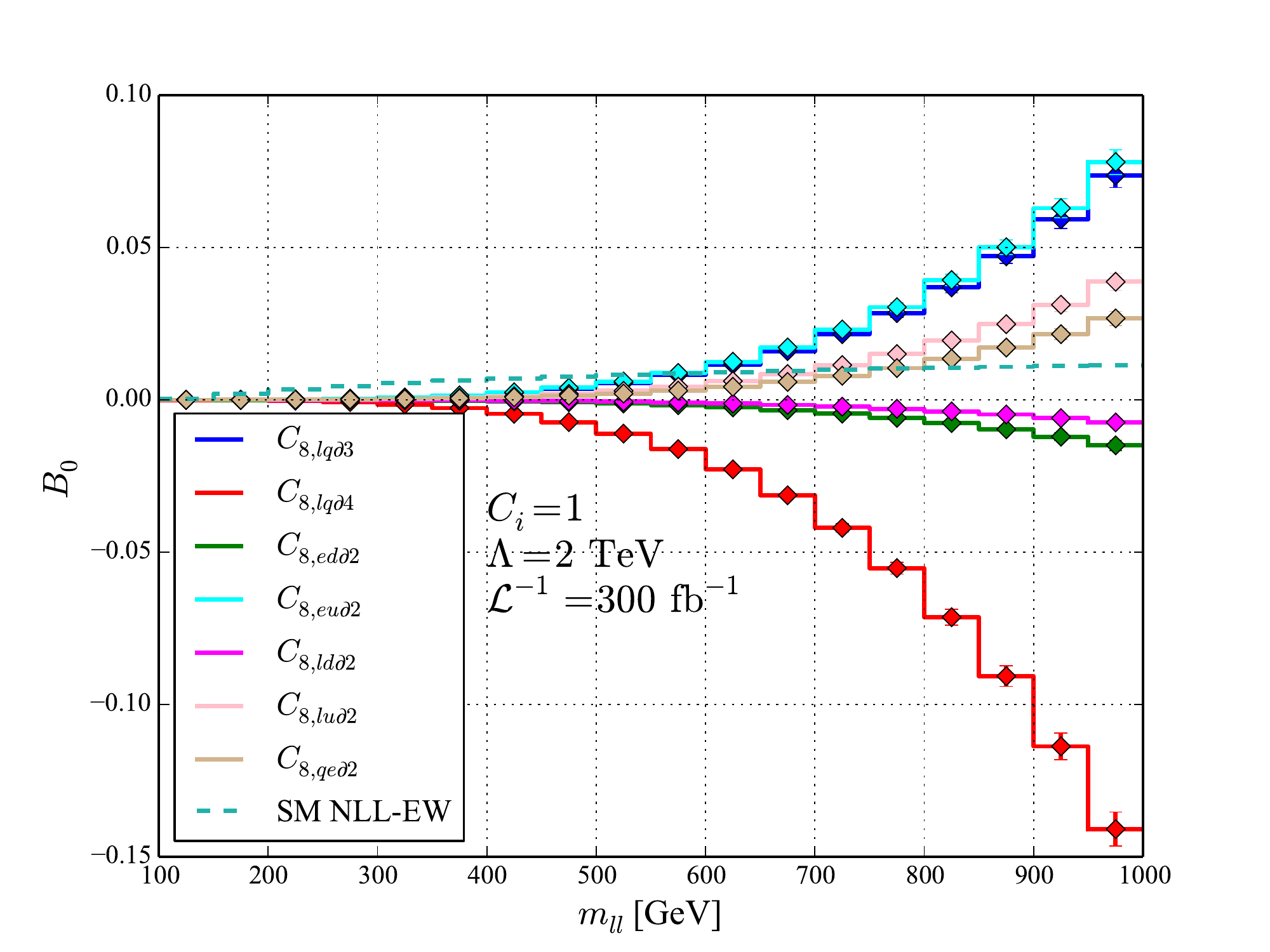}
\caption{$B_0$ coefficient as a function of the dilepton invariant mass.
\label{fig:B0}}
\end{figure}

We show in Figs.~\ref{fig:B0} numerical results for $B_0$ as a function
of the invariant mass $m_{ll}$ for the seven contributing operators.  We
set $\Lambda = 2\, {\rm TeV}$  and each
Wilson coefficient separately to $C_{i} = 1$ while
setting the others to zero to obtain these seven curves.
Although the allowed values of these coefficients have not been
determined, the value of the energy scale
$\Lambda=2$ TeV suggested by this choice is consistent with values
allowed for dimension-6 four-fermion operators found in global
fits~\cite{Berthier:2016tkq}.  We stop our plots at $m_{ll}=1$ TeV to
have a convergent EFT expansion.  The error bars denote the estimated
statistical errors for 300 fb$^{-1}$ of integrated luminosity.  We see
from Fig.~\ref{fig:B0} that searches for the $B_0$ coefficient are
promising.  A non-vanishing Wilson coefficient could be visible above
the statistical errors, while the SM contribution is small and grows
only logarithmically as opposed to polynomially like the SMEFT
effect.  We have verified that the operator ${\cal O}_{8,lq\partial 4}$ induces similarly large effects in charged-current Drell-Yan.

The results in Fig.~\ref{fig:B0} have been obtained without applying selection cuts on the final-state leptons.
Cuts on the individual lepton transverse momenta and rapidities distort the shapes of the $\theta$ and $\phi$ distributions,
so that they cannot be described in terms of Eqs.~\eqref{eq:oldCSexp} or~\eqref{eq:newCSexp}.
In standard analyses of the $A_i$ coefficients, the issue is addressed by
generating templates for the polynomials in $c_\theta$, $s_\theta$,
$c_\phi$, $s_\phi$ appearing in
Eq.~\eqref{eq:oldCSexp}~\cite{Aad:2016izn,Khachatryan:2015paa}.  A
similar strategy generalized to include the third-order polynomials in
Eq.~\eqref{eq:newCSexp} must be pursued to obtain the $B_i$ in the
presence of lepton cuts.

%%%%%%%%%%%%%%%%%%%%%%%%%%%%%%%%%%%%%%%%%%%
\section{Conclusions} \label{sec:conc}
%%%%%%%%%%%%%%%%%%%%%%%%%%%%%%%%%%%%%%%%%%%

In this note we have studied the effects of dimension-8 operators in
the SMEFT on Drell-Yan lepton-pair production at the LHC.  We have
tabulated all operators that can contribute to this process at both LO
and NLO in the QCD coupling constant.  A new angular dependence
appears associated with a class of two-derivative dimension-8
operators that is not accounted for in current studies.  Due to its
angular-momentum structure it does not appear in the SM nor in the
dimension-6 truncation of the SMEFT to any order in the QCD
perturbative expansion.  It can only be generated at higher orders by
diagrammatic contributions that connect the initial-state partons with
the final-state leptons, such as electroweak corrections.  We have shown here that these effects are
small in the SM.  To capture these new dimension-8 SMEFT effects we have proposed an extension of the
usual angular basis used when analyzing lepton pair production.  We
have demonstrated that for allowed values of the dimension-8 Wilson
coefficients that these effects would be visible at the LHC over
statistical errors.  We urge the experimental collaborations to revisit this analysis in
order to search for this clean and new signature of dimension-8 new physics.

 %%%%%%%%%%%%%%%%%%%%%%%%%%%%%%%%%%%%%%
\section{Acknowledgments}
%%%%%%%%%%%%%%%%%%%%%%%%%%%%%%%%%%%%%%

S.~A. is supported by the ERC
Starting Grant REINVENT-714788 and by the
Fondazione Cariplo and Regione Lombardia grant 2017-2070.
R.~B. is supported by the DOE contract DE-AC02-06CH11357.  E.~M. is supported by the DOE grant DE-AC52-06NA25396.
F.~P. is supported by the DOE grants DE-FG02-91ER40684 and DE-AC02-06CH11357.

\end{document}